\documentclass{llncs}
\usepackage{amsmath,amsfonts}
\usepackage{algorithm}
\usepackage{array}
\usepackage[caption=false,font=footnotesize]{subfig}
\usepackage{textcomp}
\usepackage{stfloats}
\usepackage{url}
\usepackage{verbatim}
\usepackage{graphicx}
\usepackage{cite}
\usepackage{algpseudocode}
\usepackage{float}
\newfloat{algorithm}{t}{lop}

\usepackage[dvipsnames]{xcolor}
\usepackage{listings}
\usepackage{tikz}
\usetikzlibrary{positioning}
\usetikzlibrary{calc}
\usetikzlibrary{patterns}

\usepackage{pgfplots}
\pgfplotsset{compat=newest}

\newcommand{\riscv}{\mbox{RISC-V}}

\newcommand{\fencet}{\mbox{\texttt{fence.t}}}

\newcommand{\fencets}{\mbox{\texttt{fence.t.s}}}

\newcommand{\ffclr}{\mbox{\texttt{ff.clr}}}

\usepackage[acronym]{glossaries}
\newacronym{asic}{ASIC}{Application-Specific Integrated Circuit}
\newacronym{cpu}{CPU}{central processing unit}
\newacronym{isa}{ISA}{instruction set architecture}
\newacronym[shortplural={OSes}]{os}{OS}{operating system}
\newacronym{fp}{FP}{floating-point}
\newacronym{fpu}{FPU}{floating-point unit}
\newacronym{axi}{AXI}{Advanced eXtensible Interface}
\newacronym[firstplural=systems on chip (SoCs)]{soc}{SoC}{system on chip}
\newacronym{hwpe}{HWPE}{hardware processing element}
\newacronym[firstplural=multiprocessor systems on chip (MPSoCs)]{mpsoc}{MPSoC}{multiprocessor system on a chip}
\newacronym{mpam}{MPAM}{memory partitioning and monitoring}
\newacronym{mba}{MBA}{memory bandwidth allocation}
\newacronym{qos}{QoS}{quality of service}
\newacronym{wcet}{WCET}{worst-case execution time}
\newacronym[firstplural=networks on chip (NoCs)]{noc}{NoC}{network on chip}
\newacronym{rpc}{RPC}{reduced pin count}
\newacronym{tlb}{TLB}{translation lookaside buffer}
\newacronym{bht}{BHT}{branch history table}
\newacronym{btb}{BTB}{branch target buffer}
\newacronym{lru}{LRU}{least-recently-used}
\newacronym{plru}{pseudo-LRU}{pseudo-least-recently-used}
\newacronym{lfsr}{LFSR}{linear-feedback shift register}
\newacronym{clint}{CLINT}{core-local interrupt controller}
\newacronym{csr}{CSR}{control and status register}
\newacronym{fsm}{FSM}{finite-state machine}
\newacronym{ooo}{OoO}{out-of-order}
\newacronym{rat}{RAT}{register allocation table}
\newacronym{l1d}{L1D}{L1 data cache}
\newacronym{l1i}{L1I}{L1 instruction cache}
\newacronym[firstplural=static random-access memories (SRAMs)]{sram}{SRAM}{static random-access memory}
\newacronym{sw}{SW}{software}
\newacronym{hw}{HW}{hardware}

\newif \ifDraft \Draftfalse
\newif \ifBlind \Blindtrue

\ifDraft
  \newcommand{\reviewadd}[1]{{\leavevmode\color{Green!100}#1}}
  
\else
  \newcommand{\reviewadd}[1]{#1}

\fi

\usepackage{hyperref}

\begin{document}

\setlength{\emergencystretch}{\hsize}
\setlength{\textfloatsep}{.97em}

\title{fence.t.s: Closing Timing Channels in High-Performance Out-of-Order Cores through ISA-Supported Temporal Partitioning}

\author{Nils Wistoff\inst{1} \and Gernot Heiser\inst{2} \and Luca Benini\inst{1,3}}
\institute{IIS, ETH Z\"urich, Switzerland / Email: \email{\{nwistoff, lbenini\}@iis.ee.ethz.ch} \and UNSW Sydney, Australia / Email: \email{gernot@unsw.edu.au} \and DEI, University of Bologna, Italy}

\maketitle

\begin{abstract}
  Microarchitectural timing channels exploit information leakage between security domains that should be isolated, bypassing the operating system's security boundaries.
  These channels result from contention for shared microarchitectural state.
  In the \riscv{} instruction set, the temporal fence instruction (fence.t) was proposed to close timing channels by providing an operating system with the means to temporally partition microarchitectural state inexpensively in simple in-order cores.
  This work explores challenges with fence.t in superscalar out-of-order cores featuring large and pervasive microarchitectural state.
  To overcome these challenges, we propose a novel SW-supported temporal fence (fence.t.s), which reuses existing mechanisms and supports advanced microarchitectural features, enabling full timing channel protection of an exemplary out-of-order core (OpenC910) at negligible hardware costs and a minimal performance impact of 1.0\,\%.
\end{abstract}


\section{Introduction and Related Work}

Computing systems are often time-shared between mutually untrusting applications, such as servers running different users' workloads, cyber-physical systems running control and service tasks concurrently, and personal devices that process sensitive data while accessing the internet.
To prevent malicious attackers and faulty programs from accessing restricted data, modern \glspl{isa} specify a set of mechanisms for memory protection, allowing an \gls{os} to isolate the memory views of concurrent applications.

However, as the Spectre attacks~\cite{Kocher2018spectre} have demonstrated, existing memory protection is insufficient to isolate applications reliably.
Microarchitectural \emph{timing channels} create undesired information transfer across security boundaries.
They leverage observable differences in an application's execution time depending on the microarchitectural resource usage of a previously running application.

Some works propose spatial partitioning~\cite{Page2005Partition,Domnitser2012NoMo,Kiriansky2018dawg} or randomisation~\cite{Wang2007DynamicCache1,Qureshi2018CEASER,Werner2019ScatterCache} to close timing channels in data caches.
However, these approaches do not address other microarchitectural components, such as branch predictors or buffers.
As a general solution, Ge et al.\ propose \emph{time protection}: A methodology for the \gls{os} to partition time-shared \gls{hw}~\cite{Ge2018,Ge2019a}.
However, they observe that most processors do not provide the means to temporally partition (i.e.,\ flush) on-core microarchitectural state and call for an \gls{isa} extension.
Wistoff et al.\ propose the temporal fence instruction, \fencet{}, to systematically clear microarchitectural state and thus allow the \gls{os} to fully isolate applications~\cite{wistoff2021date,wistoff2023tcomp}.
A similar approach was presented by Escouteloup et al.~\cite{Escouteloup2021Dome}.
While \fencet{} was shown to be highly effective and inexpensive on a simple in-order core, the viability of \fencet{} on complex \gls{ooo} cores remains unknown.

Evaluating \fencet{} in complex \gls{ooo} cores is relevant for multiple reasons:
(a) \gls{ooo} cores are commonly deployed in security-critical environments such as those described above,
(b) they contain large and often pervasive microarchitectural state that needs to be partitioned, 
(c) flushing and restoring on-core state to and from off-core memory incurs higher penalties and can, therefore, be expected to be more expensive than in simple cores,
(d) the overall \gls{hw}-\gls{sw} implications of dynamically clearing complex microarchitectural state have not been studied yet.
Hence, we specifically make the following contributions:

\begin{itemize}
  \item we show that integrating \fencet{} in an \gls{ooo} core with advanced microarchitectural features combining architectural and microarchitectural state comes with challenges,
  \item we address these challenges by presenting the \gls{sw}-supported temporal fence (\fencets{}) methodology, 
  \item we implement \fencets{} in a fully-featured, open-sourced, commercial, high-performance, 64-bit \riscv{} core, namely OpenC910,
  \item we show that \fencets{} closes all on-core timing channels in OpenC910 without measurable \gls{hw} overhead at a minimal performance impact of 1.0\,\%.
\end{itemize}

\section{Background}

\subsection{Processor State}
\label{s:state}

In the remainder of this paper, we will use the term \emph{architectural state} to refer to all processor state that is specified as part of the \gls{isa} and thus can be accessed directly by \gls{sw}.
Examples are the logical registers, \glspl{csr}, and memory.
Conversely, \emph{microarchitectural state} refers to all state that is transparent and not directly accessible by \gls{sw}, such as caches, branch predictors, and potentially any state-holding elements within the core, such as flip-flops and \glspl{sram} in the processor pipeline.

\subsection{Timing Channels}

A \emph{covert channel} is a communication channel that uses mechanisms that are not intended for information transfer~\cite{Lampson_73}.
A microarchitectural \emph{timing channel} is a covert channel that leverages differences in execution time due to contention for shared microarchitectural resources.
Exploitable resources are any microarchitectural, sequential components with a possible timing impact, such as data caches~\cite{Hu_92,percival2005cache}, instruction caches~\cite{Aciiccmez2007ICache}, and branch predictors~\cite{Aciiccmez2007BP}.


\subsection{Time Protection and fence.t}
\label{s:fencet}

Ge et al.\ have proposed time protection as a methodology to prevent timing channels~\cite{Ge2019a}.
The key idea is to consistently partition all shared \gls{hw} resources, either spatially (e.g.,\ through \emph{colouring} the last level cache~\cite{kessler1992colouring}) or temporally (i.e.,\ resetting the resource to a pre-defined value).

The temporal fence instruction, \fencet{}, was proposed to temporally partition on-core microarchitectural state~\cite{wistoff2021date,wistoff2023tcomp}.
It achieves this by writing back dirty cache state, clearing on-core \glspl{sram} and microarchitectural flip-flops, and padding its completion to a worst-case execution time.

\subsection{OpenC910}
\label{s:background-c910}

OpenC910 is an industrial, 64-bit, application-class, 12-stage, \gls{ooo} \riscv{} core developed by T-Head Semiconductor Co., Ltd., featuring the RV64GCXtheadc \gls{isa}~\cite{Chen2020c910}.
It was open-sourced in 2021 under the Apache license and features the custom Xtheadc extension~\cite{THead_2021c910}.

In this work, we leverage the following custom instructions and \glspl{csr}:
\texttt{dcache.call} clears the \gls{l1d}, writing back any dirty cache lines.
The \texttt{mcor} \gls{csr} enables targeted invalidations of different \glspl{sram} in the design.
\gls{sw} can trigger an invalidation of the L1 caches and branch predictors by setting corresponding bits.
\texttt{sync.i} serves as an execution barrier in the instruction stream, ensuring that all instructions preceding it have been completed before executing instructions following it.
The \texttt{mrvbr} \gls{csr} contains the address from which the core will start executing after reset.

\section{Temporally Partitioning Out-of-Order Cores}


\subsection{Challenges of fence.t in Complex Cores}
\label{s:limitations}

\paragraph{Challenge 1: The Problem of Mixed State}

\fencet{} (\autoref{s:fencet}) clears all microarchitectural state and retains architectural state as defined in \autoref{s:state}, meaning that while clearing the processor's microarchitectural state, functionally, \fencet{} behaves as a \texttt{nop}.

This approach requires a clear separation of architectural and microarchitectural state, which may not be possible in a complex processor.
For instance, \emph{register renaming}\reviewadd{, a mechanism previously targeted in side-channel attacks~\cite{May2001Renaming},} dynamically allocates logical registers to a large physical register file to mitigate data dependencies.
A \gls{rat} stores this mapping between logical and physical registers.
The \gls{rat} is not specified in the \gls{isa} or accessible by \gls{sw}, classifying it as microarchitectural state.
Since it impacts execution time, it can be exploited as a timing channel and, therefore, needs to be cleared by \fencet{}.
However, it also contains crucial information on the location of the processor's logical registers.
Simply clearing it on \fencet{} would, therefore, corrupt the processor's architectural state.
We refer to such state as \emph{mixed} state, since it combines microarchitectural and architectural information.
Handling mixed state in a way that does not leak timing information while maintaining functional correctness is challenging to achieve in \gls{hw} alone. \autoref{s:fencets} will propose \gls{sw} support to achieve these goals.

\paragraph{Challenge 2: Reusability of Instructions}
The \fencet{} instruction performs multiple operations, such as clearing SRAMs, flip-flops, and padding for a worst-case execution time.
These operations are orchestrated by a \gls{fsm} in \gls{hw} upon executing \fencet{}.
However, this monolithic approach does not give the \gls{os} any flexibility in selecting mitigation mechanisms according to the system's security and performance requirements, nor does it allow the reuse of the mechanisms for other purposes.
Therefore, we propose to split the functionality of \fencet{} in multiple instructions performing single tasks.

\subsection{Software-Supported Temporal Fence}
\label{s:fencets}

To address the challenges introduced in \autoref{s:limitations}, we propose a \gls{sw}-supported temporal fence (\fencets{}) by splitting the functionality of \fencet{} into discrete instructions and executing these as required.
An overview of this approach is shown in \autoref{a:fencets}.

\begin{algorithm}
  \caption{\fencets{} procedure.}\label{a:fencets}
  \begin{minipage}{.35\linewidth}
  \begin{algorithmic}[1]
    \ForAll{$r \in ArchRegs$}
      \State stack $ \gets r$
    \EndFor
    \State $scratch \gets sp$
  \end{algorithmic}
  \end{minipage}
  \hfill
  \begin{minipage}{.25\linewidth}
  \begin{algorithmic}[1]
    \setcounter{ALG@line}{4}
    \State \Call{ClearL1D}{}
    \State \Call{InvalSRAMs}{}
    \State \Call{ClearFFs}{}
    \State $sp \gets scratch$
  \end{algorithmic}
  \end{minipage}
  \hfill
  \begin{minipage}{.35\linewidth}
  \begin{algorithmic}[1]
    \setcounter{ALG@line}{8}
    \ForAll{$r \in ArchRegs$}
      \State $r \gets$ stack
    \EndFor
    \State \Call{PadTime}{}
  \end{algorithmic}
  \end{minipage}
\end{algorithm}

First, all architectural (logical) registers are stored on the stack.
The stack pointer is saved at a well-known location not affected by the temporal fence, such as a scratch \gls{csr}.
Next, the \gls{l1d} is cleared by writing all dirty entries back into higher-level memory.
This is followed by a (set of) instruction(s) that invalidate on-core \glspl{sram} such as caches, TLBs, and branch predictors.
Once this is done, we clear all on-core flip-flops except for the \glspl{csr} and resume execution at the next instruction.
We restore the stack pointer and the architectural registers and finally pad the execution time to a worst-case delay to prevent leakage through the context-switch latency.
This worst-case delay happens when the microarchitecture is not trained for kernel execution, and the \gls{l1d} is fully dirty and needs to be written back completely in line 5 of \autoref{a:fencets}.
In particular, note that by saving the logical registers on the stack, the \gls{rat} can be safely cleared in line 7 of \autoref{a:fencets} without losing the architectural register values, solving the problem of mixed state discussed in \autoref{s:fencets}.

\section{Implementing fence.t.s in OpenC910}
\label{s:c910}

We implement \fencets{} in the industrial, 12-stage, \gls{ooo} OpenC910 core introduced in \autoref{s:background-c910}.
We select this core as an evaluation target due to its openness, high performance, and commercial background, making it a suitable target for a representative case study.

The only \gls{hw} modification needed in OpenC910 is the addition of the \ffclr{} instruction, which clears all on-core flip-flops except for the \glspl{csr} according to line 7 of \autoref{a:fencets}.
To do so, we extend the decoder and the synchronous reset controller and route the required pipeline signals.

\begin{figure}[t]
\begin{lstlisting}[caption={\fencets{} in OpenC910.}, label={l:fencets}, escapeinside=||]
|\includegraphics[width=\linewidth]{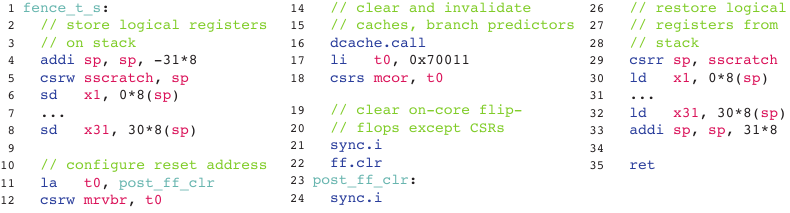}|
\end{lstlisting}
\vspace{-1em}
\end{figure}

\autoref{l:fencets} shows an implementation of \fencets{} in OpenC910.
We start by storing the logical registers on the stack.
The stack pointer is saved on the standard \riscv{} \texttt{sscratch} \gls{csr}. 
Next, we load the address of the instruction after \ffclr{} into the custom \texttt{mrvbr} \gls{csr} to continue execution from this address after coming out of reset.
Then, the data cache is cleared with \texttt{dcache.call}, and the OpenC910's on-core \glspl{sram} (caches, branch predictors) are invalidated using the custom \texttt{mcor} \gls{csr}.
Now, we can safely execute \ffclr{} to clear all on-core flip-flops.
We insert \texttt{sync.i} instructions to prevent a premature reset.
Finally, we restore the logical registers from the stack and return from the function.
We pad the execution time of \fencets{} to 15\,k cycles, which proved sufficient in all experiments even with a fully dirty \gls{l1d}.

\section{Evaluation}
\subsection{Platform}

For our evaluation, we embed OpenC910 into a minimal system with peripherals and memory required to preload and execute binaries.
We map this design onto a Xilinx VCU128 FPGA board running at 50\,MHz.
\autoref{f:vcu128} illustrates this setup.

Furthermore, we port an experimental version of seL4, a formally verified microkernel with strong security guarantees~\cite{Klein2014_seL4}, onto this system.
Since this work focuses on on-core timing channels (within \emph{Core 0} shown in \autoref{f:vcu128}), we configure seL4 to colour the L2 cache of OpenC910 to prevent timing interference through off-core cache refill latencies~\cite{kessler1992colouring}.

\newlength\figH
\newlength\figW
\setlength{\figW}{0.38\linewidth}
\setlength{\figH}{.7\figW}
\def\cmScale{0.72}

\pgfplotsset{cmStyle/.style={
    x tick label style={/pgf/number format/.cd, precision=1},
    scaled y ticks = false,
    y tick label style={/pgf/number format/.cd, fixed, precision=0},
    xtick distance = 128,
    colorbar style = {at={(1.05,0)}, anchor=south west, scaled y ticks=false}
  }}

\captionsetup[subfloat]{captionskip=0pt}

\begin{figure}[t]
  \centering
    \begin{minipage}[b]{0.49\linewidth}
      \centering
      \includegraphics[width=\linewidth]{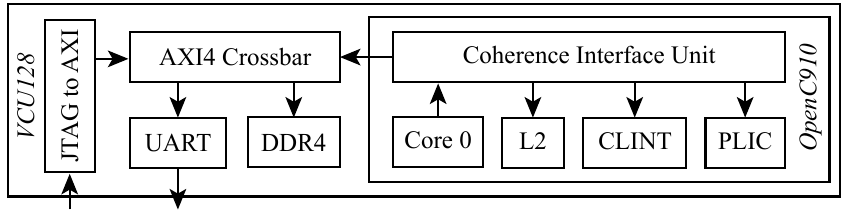}
      \vspace{-1ex}
      \caption{FPGA evaluation platform.}
      \label{f:vcu128}
    \end{minipage}
    \hfill
    \begin{minipage}[b]{0.5\linewidth}
      \centering
      \begin{tikzpicture}
    \begin{axis}[
        width  = 1.09\textwidth,
        height = 2.5cm,
        major x tick style = transparent,
        ybar=2*\pgflinewidth,
        bar width=6pt,
        ymajorgrids = true,
        minor y tick num = 1,
        yminorgrids = true,
        ylabel = {\scriptsize Slowdown},
        ylabel style={yshift=-2pt, xshift=-6pt},
        symbolic x coords={barnes,cholesky,fft,fmm,lu,ocean,radiosity,radix,raytrace,watern-\\[-1ex]squared,water-\\[-1ex]spatial},
        xtick = data,
        xtick align=outside,
        xtick style={},
        xticklabel style={font=\scriptsize,rotate=50,anchor=east,xshift=2,yshift=-3pt,align=right},
        scaled y ticks = false,
        ytick distance=1,
        yticklabel={\scriptsize\pgfmathprintnumber\tick\%},
        enlarge x limits=0.05,
        ymin=0
    ]
        \addplot[fill=black,style={mark=none}]
            coordinates {(barnes, 1.20) (cholesky, 0.53) (fft, 1.05) (fmm, 1.25) (lu, 0.82) (ocean, 0.45) (radiosity, 1.57) (radix, 0.68) (raytrace, 0.85) (watern-\\[-1ex]squared, 1.33) (water-\\[-1ex]spatial, 1.52)};

    \end{axis}
\end{tikzpicture}
      \vspace{-2.8em}
      \caption{Splash-2 slowdown by \fencets{}.}
      \label{f:bm}
    \end{minipage}
    \
    \
    \begin{minipage}[b]{\linewidth}
      \centering
      \subfloat[L1D. Unmitigated.]{%
          \begin{minipage}[b]{0.31\linewidth}
              \centering
              \footnotesize
\begin{tikzpicture}[scale=\cmScale]

\definecolor{color0}{rgb}{0.267004,0.004874,0.329415}

\begin{axis}[cmStyle,
axis background/.style={fill=color0},
colormap/viridis,
height=\figH,
point meta max=0.2916014790535,
point meta min=0,
tick align=outside,
tick pos=left,
width=\figW,
x grid style={white!69.0196078431373!black},
xlabel={Secret},
xmin=0, xmax=513,
xtick style={color=black},
y grid style={white!69.0196078431373!black},
ylabel={Time (cycles)},
ymin=41785, ymax=51003,
ytick style={color=black}
]
\addplot graphics [includegraphics cmd=\pgfimage,xmin=0, xmax=513, ymin=41785, ymax=51003] {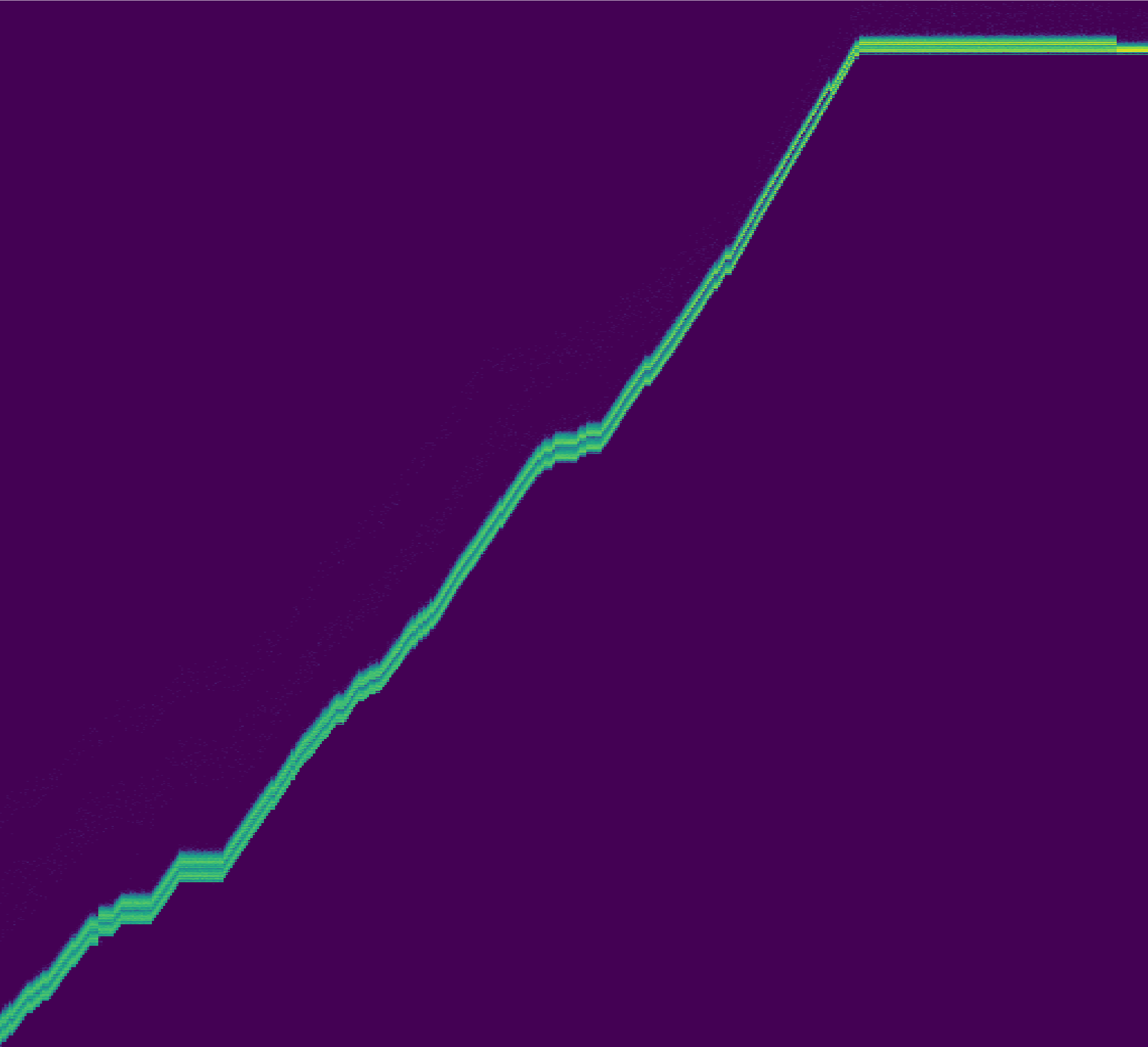};
\end{axis}

\end{tikzpicture}
              \vspace{0.5em}
              \scriptsize
              $\mathcal{M} = 4283\,\text{mb}$,\,$\mathcal{M}_0 = 0.7\,\text{mb}$
          \end{minipage}
          \label{f:l1d-unmitigated}%
      }
      \hfill
      \subfloat[L1I. Unmitigated.]{%
          \begin{minipage}[b]{0.28\linewidth}
              \centering
              \footnotesize
\begin{tikzpicture}[scale=\cmScale]

\definecolor{color0}{rgb}{0.267004,0.004874,0.329415}

\begin{axis}[cmStyle,
axis background/.style={fill=color0},
colormap/viridis,
height=\figH,
point meta max=0.5197095870972,
point meta min=0,
tick align=outside,
tick pos=left,
width=\figW,
x grid style={white!69.0196078431373!black},
xlabel={Secret},
xmin=0, xmax=513,
xtick style={color=black},
y grid style={white!69.0196078431373!black},
ymin=41547, ymax=72248,
ytick style={color=black}
]
\addplot graphics [includegraphics cmd=\pgfimage,xmin=0, xmax=513, ymin=41547, ymax=72248] {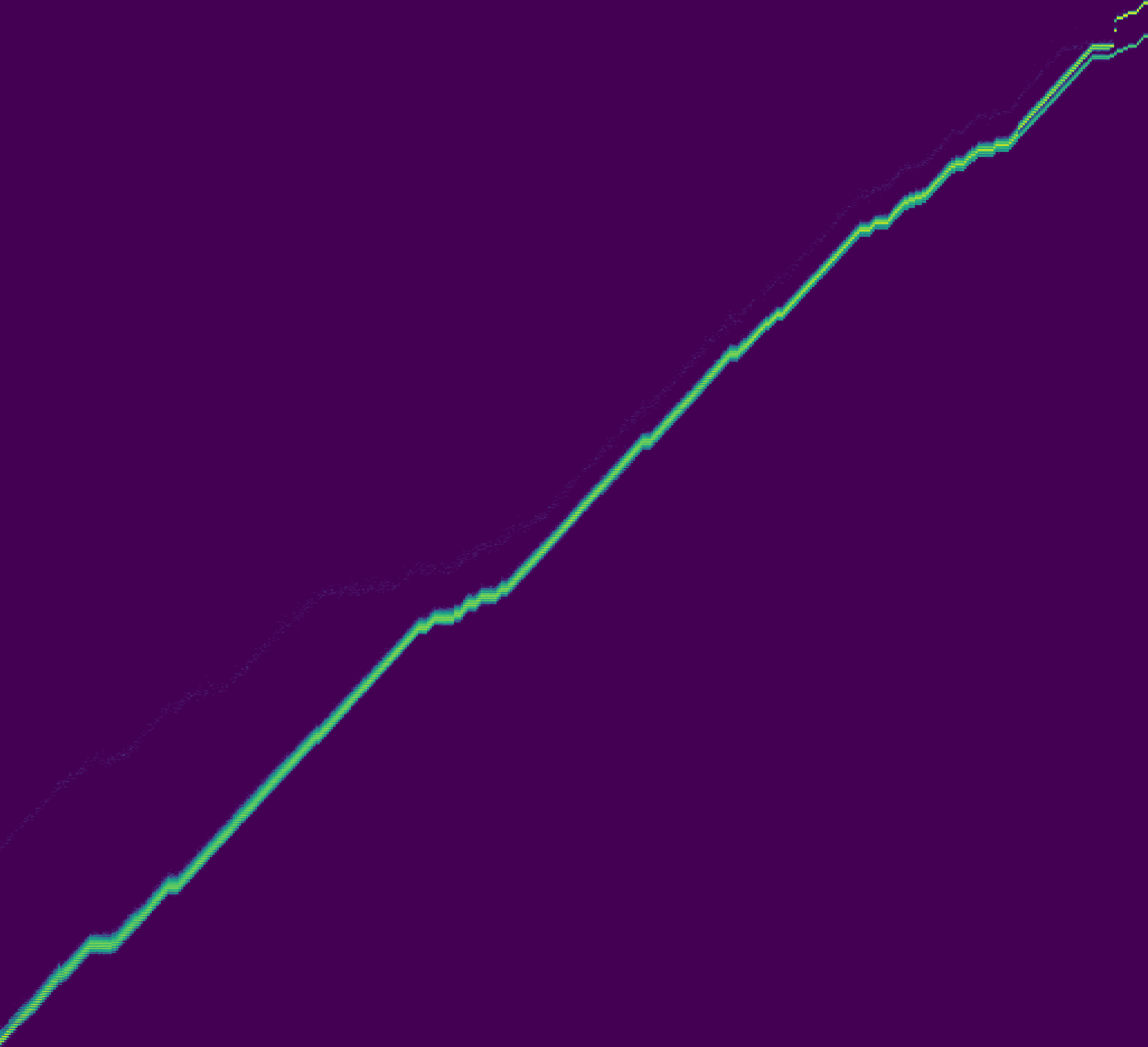};
\end{axis}

\end{tikzpicture}
              \vspace{0.5em}
              \scriptsize
              $\mathcal{M} = 5940\,\text{mb}$,$\mathcal{M}_0 = 0.8\,\text{mb}$
          \end{minipage}
          \label{f:l1i-unmitigated}%
      }
      \hfill
      \pgfplotsset{cmStyle/.style={
          x tick label style={/pgf/number format/.cd, precision=1},
          scaled y ticks = false,
          y tick label style={/pgf/number format/.cd, fixed, precision=0},
          xtick distance = 32,
          colorbar style = {at={(1.05,0)}, anchor=south west, scaled y ticks=false}
      }}
      \subfloat[BHT. Unmitigated.]{%
          \begin{minipage}[b]{0.37\linewidth}
              \centering
              \footnotesize
\begin{tikzpicture}[scale=\cmScale]

\definecolor{color0}{rgb}{0.267004,0.004874,0.329415}

\begin{axis}[cmStyle,
axis background/.style={fill=color0},
colorbar,
colorbar style={ytick={0,0.0138797819356963,0.0277595638713927,0.041639345807089,0.0555191277427853,0.0633503020430121,0.0697488373425227,0.0751587218546047,0.0798449769761886,0.0839785469422601,0.0876761512764154,0.112002000509819,0.126231710109556,0.136327849743222,0.144159024043449,0.150557559342959,0.155967443855041,0.160653698976625,0.164787268942697,0.168484873276852},yticklabels={\(\displaystyle {0}\),,,,,,,,,,\(\displaystyle {10^{-2}}\),,,,,,,,,\(\displaystyle {10^{-1}}\)},ylabel={Probability}},
colormap/viridis,
height=\figH,
point meta max=0.1912386417389,
point meta min=0,
tick align=outside,
tick pos=left,
width=\figW,
x grid style={white!69.0196078431373!black},
xlabel={Secret},
xmin=0, xmax=129,
xtick style={color=black},
y grid style={white!69.0196078431373!black},
ymin=1122, ymax=1400,
ytick style={color=black}
]
\addplot graphics [includegraphics cmd=\pgfimage,xmin=0, xmax=129, ymin=1122, ymax=1400] {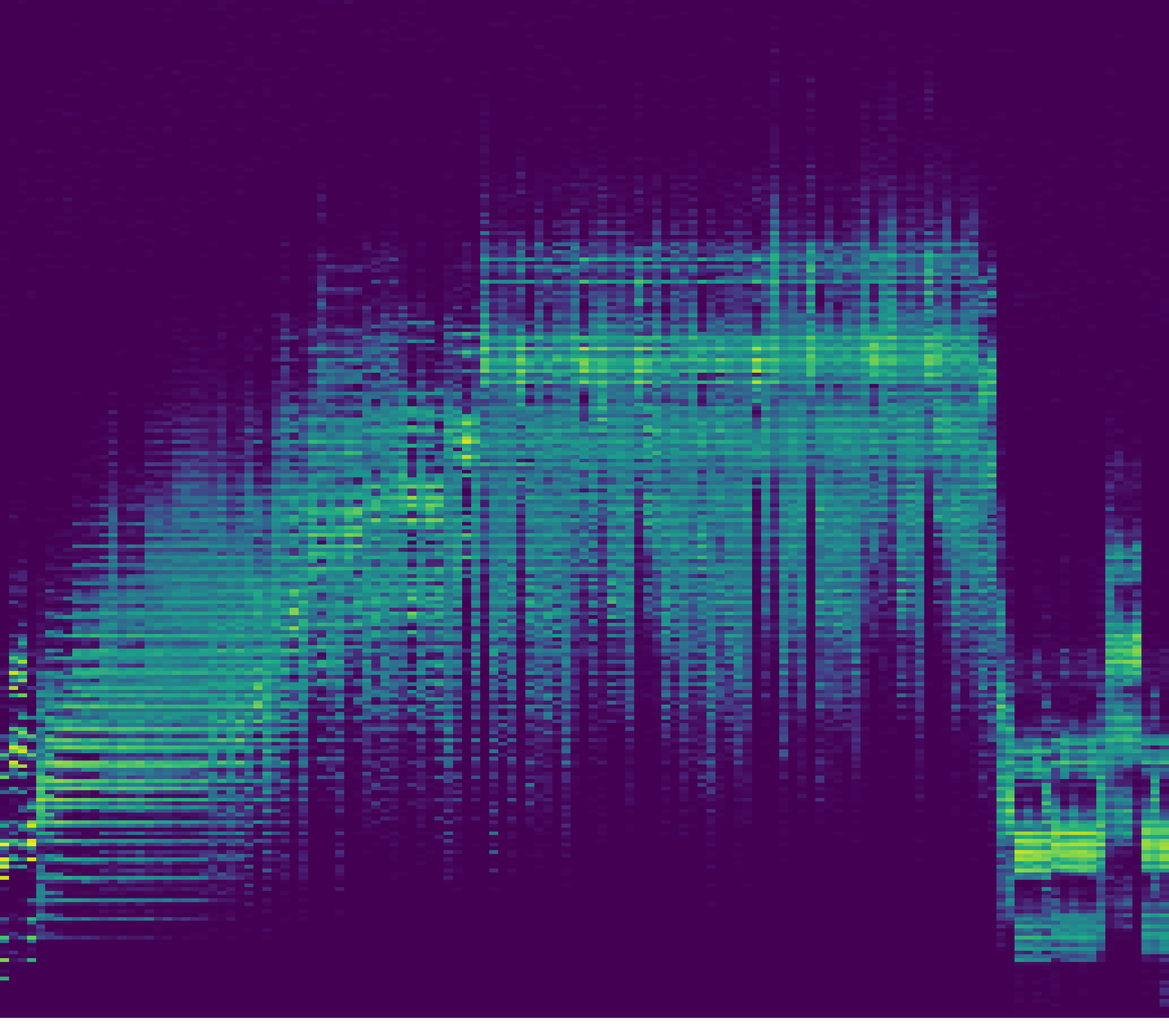};
\end{axis}

\end{tikzpicture}
              \vspace{0.5em}
              \scriptsize
              $\mathcal{M} = 698.8\,\text{mb}$, $\mathcal{M}_0 = 1.1\,\text{mb}$
          \end{minipage}
          \label{f:bht-unmitigated}%
      }
      \\[-1ex]
      \pgfplotsset{cmStyle/.style={
          x tick label style={/pgf/number format/.cd, precision=1},
          scaled y ticks = false,
          y tick label style={/pgf/number format/.cd, fixed, precision=0},
          xtick distance = 128,
          colorbar style = {at={(1.05,0)}, anchor=south west, scaled y ticks=false}
        }}
      \subfloat[L1D. \fencets{}.]{%
          \begin{minipage}[b]{0.31\linewidth}
              \centering
              \footnotesize
\begin{tikzpicture}[scale=\cmScale]

\definecolor{color0}{rgb}{0.267004,0.004874,0.329415}

\begin{axis}[cmStyle,
axis background/.style={fill=color0},
colormap/viridis,
height=\figH,
point meta max=0.9860681295395,
point meta min=0,
tick align=outside,
tick pos=left,
width=\figW,
x grid style={white!69.0196078431373!black},
xlabel={Secret},
xmin=0, xmax=513,
xtick style={color=black},
y grid style={white!69.0196078431373!black},
ylabel={Time (cycles)},
ymin=51300, ymax=51400,
ytick style={color=black}
]
\addplot graphics [includegraphics cmd=\pgfimage,xmin=0, xmax=513, ymin=51300, ymax=51400] {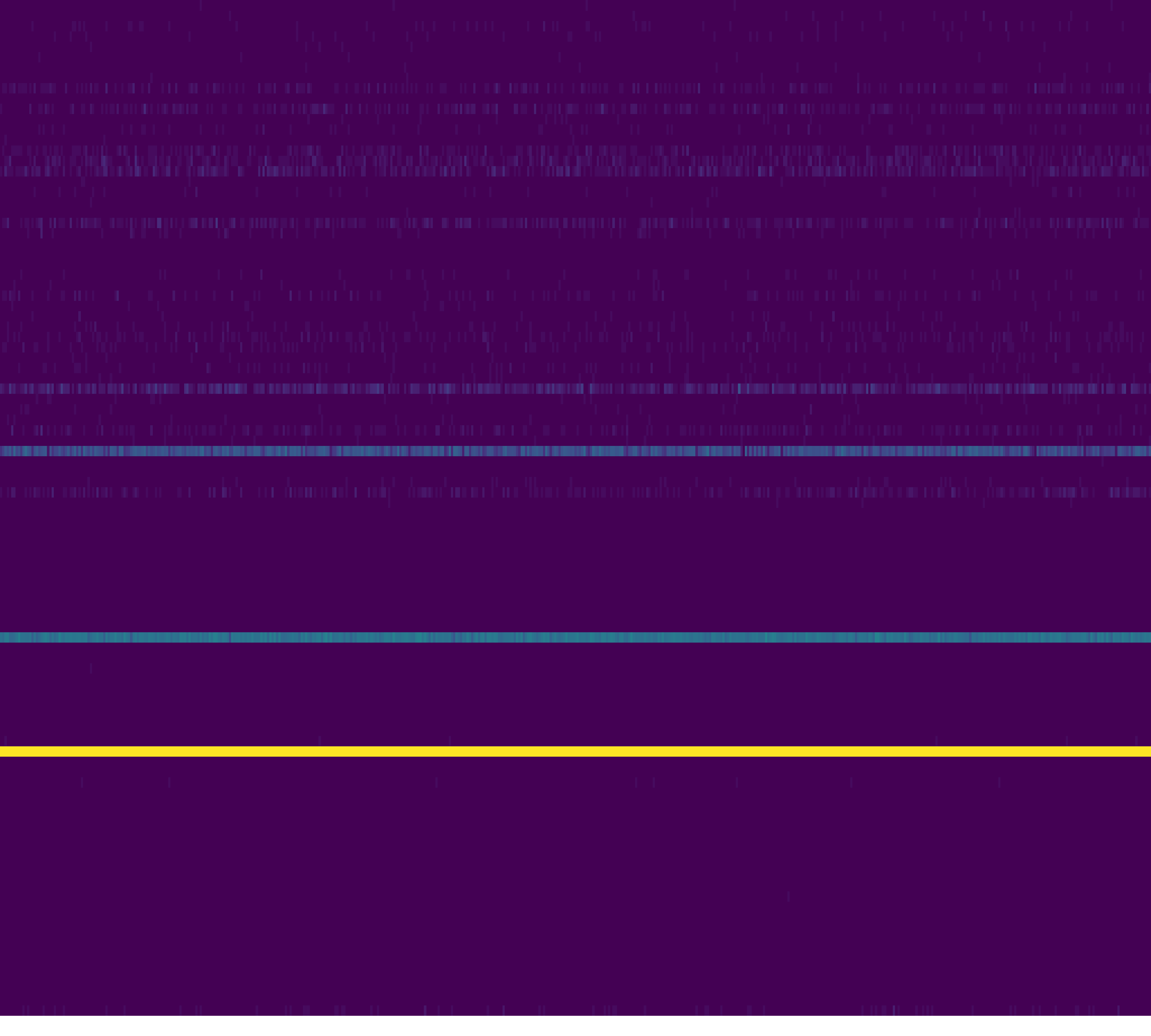};
\end{axis}

\end{tikzpicture}
              \vspace{0.5em}
              \scriptsize
              $\mathcal{M} = 64.3\,\text{mb}$, $\mathcal{M}_0 = 71.0\,\text{mb}$
          \end{minipage}
          \label{f:l1d-fencet}%
      }
      \hfill
      \subfloat[L1I. \fencets{}.]{%
          \begin{minipage}[b]{0.28\linewidth}
              \centering
              \footnotesize
\begin{tikzpicture}[scale=\cmScale]

\definecolor{color0}{rgb}{0.267004,0.004874,0.329415}

\begin{axis}[cmStyle,
axis background/.style={fill=color0},
colormap/viridis,
height=\figH,
point meta max=0.7751479148865,
point meta min=0,
tick align=outside,
tick pos=left,
width=\figW,
x grid style={white!69.0196078431373!black},
xlabel={Secret},
xmin=0, xmax=513,
xtick style={color=black},
y grid style={white!69.0196078431373!black},
ymin=70600, ymax=70800,
ytick style={color=black}
]
\addplot graphics [includegraphics cmd=\pgfimage,xmin=0, xmax=513, ymin=70600, ymax=70800] {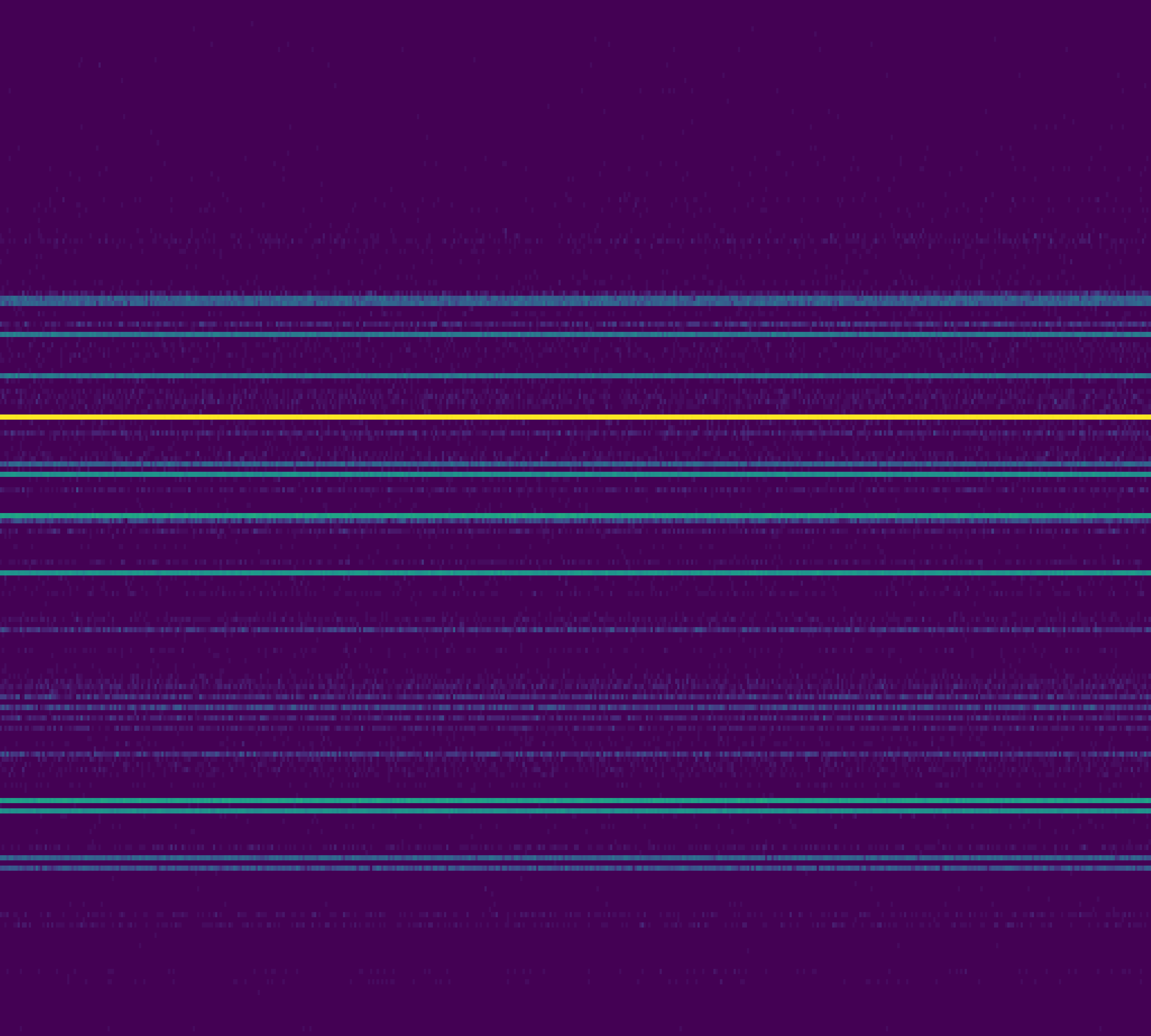};
\end{axis}

\end{tikzpicture}
              \vspace{0.5em}
              \scriptsize
              $\mathcal{M} = 3.2\,\text{mb}$,$\mathcal{M}_0 = 3.5\,\text{mb}$
          \end{minipage}
          \label{f:l1i-fencet}%
      }
      \hfill
      \pgfplotsset{cmStyle/.style={
          x tick label style={/pgf/number format/.cd, precision=1},
          scaled y ticks = false,
          y tick label style={/pgf/number format/.cd, fixed, precision=0},
          xtick distance = 32,
          colorbar style = {at={(1.05,0)}, anchor=south west, scaled y ticks=false}
      }}
      \subfloat[BHT. \fencets{}.]{%
          \begin{minipage}[b]{0.37\linewidth}
              \centering
              \footnotesize
\begin{tikzpicture}[scale=\cmScale]

\definecolor{color0}{rgb}{0.267004,0.004874,0.329415}

\begin{axis}[cmStyle,
axis background/.style={fill=color0},
colorbar,
colorbar style={ytick={0,0.0532791950387897,0.106558390077579,0.159837585116369,0.213116780155159,0.2431776748333,0.267739214183458,0.288505699809233,0.306494447887635,0.322361648211756,0.336555342565777,0.429933010298254,0.484555444326553,0.523310678030731,0.553371572708872,0.57793311205903,0.598699597684806,0.616688345763207,0.632555546087328,0.646749240441349,0.740126908173826,0.794749342202125,0.833504575906303,0.863565470584444,0.888127009934601,0.908893495560377,0.926882243638779,0.942749443962901},yticklabels={\(\displaystyle {0}\),,,,,,,,,,\(\displaystyle {10^{-2}}\),,,,,,,,,\(\displaystyle {10^{-1}}\),,,,,,,,},ylabel={Probability}},
colormap/viridis,
height=\figH,
point meta max=0.9500386118889,
point meta min=0,
tick align=outside,
tick pos=left,
width=\figW,
x grid style={white!69.0196078431373!black},
xlabel={Secret},
xmin=0, xmax=129,
xtick style={color=black},
y grid style={white!69.0196078431373!black},
ymin=4780, ymax=4820,
ytick style={color=black}
]
\addplot graphics [includegraphics cmd=\pgfimage,xmin=0, xmax=129, ymin=4780, ymax=4820] {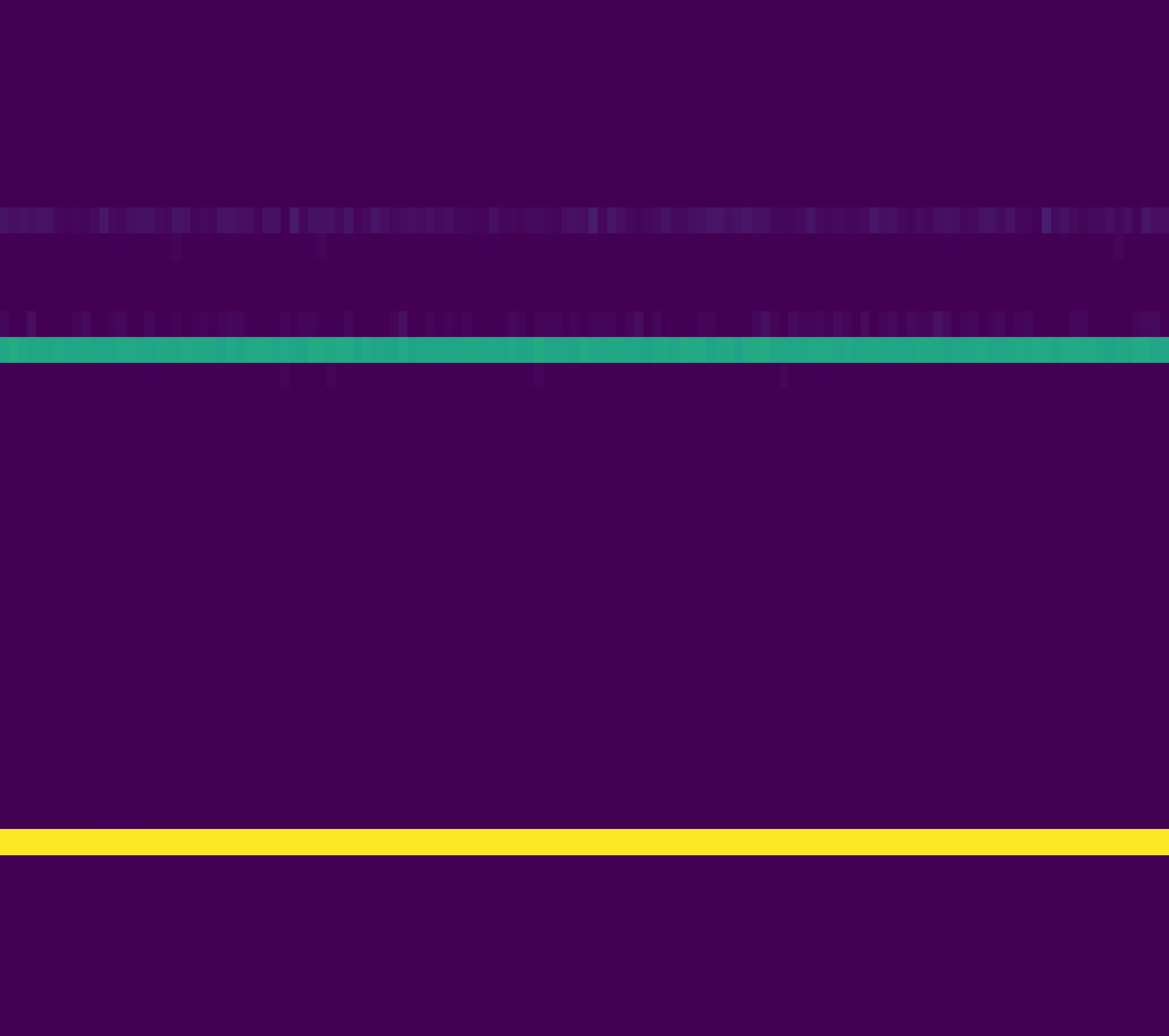};
\end{axis}

\end{tikzpicture}
              \vspace{0.5em}
              \scriptsize
              $\mathcal{M} = 3.3\,\text{mb}$, $\mathcal{M}_0 = 6.4\,\text{mb}$
          \end{minipage}
          \label{f:bht-fencet}%
      }
      \vspace{-1ex}
      \caption{Channel matrices and corresponding mutual information.}
      \label{f:cm}
    \end{minipage}
    \vspace{-3ex}
\end{figure}

\subsection{Security Analysis}

To evaluate the efficacy of \fencets{}, we adopt the methodology of previous works~\cite{Ge2018,Ge2019a,wistoff2021date,wistoff2023tcomp} and port \emph{channel bench}~\cite{Ge2018}, an evaluation framework for timing channels, to OpenC910.
\emph{Channel bench} sets up two applications, a Trojan and a spy, that actively try to communicate through different microarchitectural components using a \emph{prime-and-probe} attack~\cite{percival2005cache}.
For a given secret $s$, the Trojan performs a corresponding action (e.g. evict $s$ lines from the \gls{l1d}), and the spy subsequently measures its own execution time.
The results of these experiments are captured in \emph{channel matrices}; see \autoref{f:cm} for examples.
The x-axis displays the secret values encoded by the Trojan, and the y-axis denotes the subsequent execution time of the spy's prime function.
Colours indicate the rate of incidence for a given secret-time pair.
If a channel matrix shows a horizontal variation, this indicates a correlation between the spy's execution time and the Trojan's secret and, thus, a potential timing channel.

In addition, we use leakiEst~\cite{Chothia2013} to compute the mutual information $\mathcal{M}$ for a channel matrix to quantify the channel's capacity.
$\mathcal{M}$ measures the information gained about one random variable (e.g.,\ the Trojan's secret) by observing another (e.g.,\ the spy's execution time)~\cite{Shannon_48}.
In our security analysis, we report $\mathcal{M}$ in millibits (mb).
Since a non-zero $\mathcal{M}$ can be caused by random noise, we simulate a channel-less measurement with the same output distribution and compute the zero-leakage upper bound, $\mathcal{M}_0$, from it.
$\mathcal{M} > \mathcal{M}_0$ indicates a timing channel.

\autoref{f:cm} shows the channel bench results on OpenC910 for the \gls{l1d}, the \gls{l1i}, and the \gls{bht} without timing channel mitigations (top) and with \fencets{} on a context switch (bottom).
We use these components as examples for our security evaluation but stress that \fencets{} aims at preventing timing channels through \emph{any} microarchitectural state.
Without mitigations (top), all three components show significant leaks, evident from the diagonal patterns in their channel matrices and their $\mathcal{M}$ being several orders of magnitude greater than the corresponding $\mathcal{M}_0$.
Executing \fencets{} on a context switch (bottom) reliably closes all timing channels, as there are no variations along the x-axis.
The $\mathcal{M}$ values support this, as all three are smaller than $\mathcal{M}_0$.

\subsection{Performance Overhead}

To evaluate the performance impact of \fencets{}, we run one application executing the Splash-2 benchmarks~\cite{Woo1995splash} and one idle application concurrently on OpenC910.
seL4 performs a context switch every 10\,M cycles, corresponding to a typical time slice length of 10\,ms at a clock frequency of 1\,GHz.

\autoref{f:bm} shows the relative slowdown of the benchmarks when calling \fencets{} on a context switch.
The combined performance impact of the increased context-switch latency (direct costs) and the cold microarchitecture after context-switching (indirect costs) is, on average, 1.0\,\%, while being consistently below 1.6\,\% for all benchmark applications.

The low overhead of the temporal fence can be explained by the large number of system clock cycles per \gls{os} time slice and, as Ge et al.~\cite{Ge2018} argue, the low persistance of L1 state across multiple time slices.
We stress that the \gls{os} can choose \emph{if} and at what interval to perform the temporal fence according to the system's performance and security requirements.

\subsection{Hardware Costs}

As discussed in \autoref{s:c910}, our \gls{hw} modifications are limited to adding the minimal \ffclr{} instruction.
To evaluate the \gls{hw} costs of this modification, we synthesise the original and the extended version of OpenC910 in \textsc{GlobalFoundries} 12LP+ FinFET technology in typical corners at a clock speed of 2\,GHz.
The addition of \ffclr{} does not cause a notable change in the area or timing of the design.

\section{Conclusion}

This work proposes the \gls{sw}-supported temporal fence (\fencets{}) that closes timing channels even in complex \gls{ooo} cores.
By implementing it in OpenC910, an industry-grade, high-performance \riscv{} core, we observe that this processor already provides a set of mechanisms to clear most on-core state required for time protection.
We add the simple \ffclr{} instruction, which clears all on-core flip-flops except for thee \glspl{csr} at negligible \gls{hw} costs, and demonstrate that an \gls{os} can combine it with custom instructions already provided by OpenC910 to reliably close timing channels at a minimal performance overhead of 1.0\,\%.

We conclude that by exposing mechanisms already available in commercial \gls{hw} to \gls{sw}, an \gls{os} can efficiently enforce temporal isolation when needed.
Adding this abstraction to the \gls{isa} would enable a unified set of mechanisms to close timing channels across different \riscv{} implementations.

\section*{Acknowledgements}
This work was supported in part through the TRISTAN (101095947) project that received funding from the HORIZON CHIPS-JU programme.

\bibliographystyle{splncs}
\bibliography{refs}

\begin{thebibliography}{10}

\bibitem{Kocher2018spectre}
Kocher, P. et~al.:
\newblock Spectre attacks: Exploiting speculative execution.
\newblock In: IEEE S\&P. (2019)  1--19

\bibitem{Page2005Partition}
Page, D.:
\newblock Partitioned cache architecture as a side-channel defence mechanism.
\newblock IACR Cryptology ePrint Archive (2005)

\bibitem{Domnitser2012NoMo}
Domnitser, L. et~al.:
\newblock Non-monopolizable caches: Low-complexity mitigation of cache side
  channel attacks.
\newblock ACM TACO \textbf{8} (2012)  35:1--35:21

\bibitem{Kiriansky2018dawg}
Kiriansky, V. et~al.:
\newblock Dawg: A defense against cache timing attacks in speculative execution
  processors.
\newblock In: IEEE/ACM MICRO. (2018)  974--987

\bibitem{Wang2007DynamicCache1}
Wang, Z. et~al.:
\newblock New cache designs for thwarting software cache-based side channel
  attacks.
\newblock In: IEEE/ACM ISCA. (2007)  494--505

\bibitem{Qureshi2018CEASER}
Qureshi, M.~K.:
\newblock {CEASER}: Mitigating conflict-based cache attacks via
  encrypted-address and remapping.
\newblock IEEE/ACM MICRO (2018)  775--787

\bibitem{Werner2019ScatterCache}
Werner, M. et~al.:
\newblock {SCATTERCACHE}: Thwarting cache attacks via cache set randomization.
\newblock In: USENIX SEC. (2019)  675--692

\bibitem{Ge2018}
Ge, Q. et~al.:
\newblock No security without time protection: We need a new hardware-software
  contract.
\newblock In: ACM APSys. (2018)  1:1--1:9

\bibitem{Ge2019a}
Ge, Q. et~al.:
\newblock Time protection: The missing {OS} abstraction.
\newblock In: ACM EuroSys. (2019)  1:1--1:17

\bibitem{wistoff2021date}
Wistoff, N. et~al.:
\newblock Microarchitectural timing channels and their prevention on an
  open-source 64-bit {RISC-V} core.
\newblock In: DATE. (2021)  627--632

\bibitem{wistoff2023tcomp}
Wistoff, N. et~al.:
\newblock Systematic prevention of on-core timing channels by full temporal
  partitioning.
\newblock IEEE. Trans. Comput. \textbf{72}(5) (2023)  1420--1430

\bibitem{Escouteloup2021Dome}
Escouteloup, M. et~al.:
\newblock Under the dome: preventing hardware timing information leakage.
\newblock In: CARDIS. (2021)  1--20

\bibitem{Lampson_73}
Lampson, B.~W.:
\newblock A note on the confinement problem.
\newblock Commun. ACM \textbf{16} (1973)  613--615

\bibitem{Hu_92}
Hu, W.-M.:
\newblock Lattice scheduling and covert channels.
\newblock In: IEEE S\&P. (1992)  52--61

\bibitem{percival2005cache}
Percival, C.:
\newblock Cache missing for fun and profit.
\newblock In: BSDCan. (2005)

\bibitem{Aciiccmez2007ICache}
Ac{\i}i{\c{c}}mez, O.:
\newblock Yet another microarchitectural attack: Exploiting i-cache.
\newblock In: ACM CSAW. (2007)  11--18

\bibitem{Aciiccmez2007BP}
Ac{\i}i{\c{c}}mez, O. et~al.:
\newblock Predicting secret keys via branch prediction.
\newblock In: CT-RSA. (2007)  225--242

\bibitem{kessler1992colouring}
Kessler, R.~E. et~al.:
\newblock Page placement algorithms for large real-indexed caches.
\newblock ACM TOCS \textbf{10}(4) (1992)  338--359

\bibitem{Chen2020c910}
Chen, C. et~al.:
\newblock Xuantie-910: A commercial multi-core 12-stage pipeline out-of-order
  64-bit high performance {RISC-V} processor with vector extension : Industrial
  product.
\newblock In: ACM/IEEE ISCA. (2020)  52--64

\bibitem{THead_2021c910}
{T-Head Semiconductor Co., Ltd.}:
\newblock {OpenC910} core (2021) https://github.com/T-head-Semi/openc910.

\bibitem{May2001Renaming}
May, D. et~al.:
\newblock Random register renaming to foil {DPA}.
\newblock In: CHES. (2001)  28--38

\bibitem{Klein2014_seL4}
Klein, G. et~al.:
\newblock Comprehensive formal verification of an {OS} microkernel.
\newblock ACM TOCS \textbf{32}(1) (2014)  2:1--2:70

\bibitem{Chothia2013}
Chothia, T. et~al.:
\newblock A tool for estimating information leakage.
\newblock In: CAV. (2013)  690--695

\bibitem{Shannon_48}
Shannon, C.~E.:
\newblock A mathematical theory of communication.
\newblock Bell Labs Tech. J. \textbf{27} (1948)  379--423

\bibitem{Woo1995splash}
Woo, S.~C. et~al.:
\newblock The splash-2 programs: Characterization and methodological
  considerations.
\newblock ACM SIGARCH Comput. Archit. News \textbf{23}(2) (1995)  24--36

\end{thebibliography}

\vfill

\end{document}